\documentclass[review]{elsarticle}
\usepackage{lineno,hyperref}
\usepackage{graphicx}
\usepackage{subfigure}
\usepackage{amsmath}
\usepackage[pdftex]{color}
\usepackage[font=footnotesize,labelfont=bf]{caption}

\journal{Journal of \LaTeX~templates}

\begin{document}

\begin{frontmatter}

\title{Dislocation assisted phase separation: a phase field study}
\author{Arjun Varma R.}

\author{Prita Pant}

\author{M. P. Gururajan\corref{mycorrespondingauthor}}

\cortext[mycorrespondingauthor]{Corresponding Author: 
M. P. Gururajan}
\ead{guru.mp@iitb.ac.in}

\address{Department of Metallurgical 
Engineering and Materials Science, Indian Institute 
of Technology Bombay, Powai, Mumbai, Maharashtra 400076 INDIA}


\begin{abstract}
Defects play a key role in deciding the mechanisms and kinetics 
of phase transformations. In this paper, we show 
how dislocations influence phase separation
in alloys with miscibility gap. Specifically, depending 
on the ratio of pipe mobility to bulk mobility,
it is seen that even in a system with nominal compositions outside
the spinodal limit, spinodal phase separation is possible.
Surprisingly, phase separation through both
nucleation and growth, and spinodal decomposition, is
seen concurrently (for the case of intersecting dislocations). 
Finally, the prominent role played by dislocations in
influencing the morphology of precipitates is explored. We show that these
results agree qualitatively with recent experimental results 
in iron based systems obtained using Atom Probe Tomography (APT).
\end{abstract}

\begin{keyword}
Segregation \sep Pipe diffusion \sep Precipitate morphology \sep 
Spinodal decomposition \sep Misfit \sep Nucleation and growth 
\end{keyword}

\end{frontmatter}


\section{Introduction}

It is well known that solutes in alloys segregate to 
elastically favourable sites near dislocations
~\cite{CottrellBilby1949,CottrellJaswon1949,
Bilby1950,WolferAshkin1976} 
forming ``solute atmospheres" which hinder the 
dislocation movement. The solute segregation to 
dislocations might depend on the character 
of the dislocation and the nature of the misfit (between solute and matrix), if 
only the elastic interaction is accounted for
~\cite{Beaven1980}. Solutes with dilatational misfit 
segregate to dislocations with edge character while 
solutes with non-dilatational misfit segregate to both 
edge and screw dislocations~\cite{Cochardt1955}. 
However, if the broken bonds at the dislocation 
core are accounted for, solute segregation is 
expected at dislocations of both character
~\cite{AndersonHirthLothe2017}. 

Segregation can also lead to nucleation and 
growth along dislocations as has been predicted
~\cite{Cahn1957,Ham1959,Gomez1973}, and, 
experimentally observed -- see, for example,
~\cite{Dash1956,Beaven1980,Feng2010}. 
Recently, atom probe experiments in Fe-Mn 
system \cite{DaSilva2018} have shown that dislocations 
can also act as sites of localized spinodal 
decomposition in systems with nominal alloy 
compositions outside the spinodal limit. In fact, 
in this system, namely, Fe-Mn, there have been 
recent modelling studies which show grain boundary 
induced spinodal decomposition even when
the overall alloy composition is outside the spinodal 
limits~\cite{Kamachali2020,Mianroodi2021}.
In this paper, we explore if dislocations can 
also induce phase separation by the spinodal mechanism 
in systems with nominal alloy compositions outside 
the spinodal limit.

In the past, attempts have been made to investigate 
the effects of dislocations on phase transformation 
using phase field models~\cite{Leonard1998,Hu2001,
Haataja2004,Haataja2005,Ma2006,Li2009}. These models 
have shown that stationary and moving dislocations, 
in systems with average composition within the 
spinodal limit, can lead to faster kinetics and 
ordered microstructures. In this regard, the work 
of Hu and Chen~\cite{Hu2001} and Mianroodi et al~\cite{Mianroodi2021}
deserve special mention; both these works show
that even in a system with overall composition outside 
the spinodal limit, segregation to the dislocations can take the 
composition along the dislocation to within the spinodal regime. 
However, in both these cases spinodal fluctuations along the dislocation
line are not seen. This is in  contrast to the spinodal decomposition along the 
dislocations  observed experimentally by Da Silva 
et al.~\cite{DaSilva2018}. 

The work of Hu and Chen~\cite{Hu2001} is in 2-D. So,
the question of spinodal fluctuations along the dislocation line
does not arise. On the other hand, in Mianroodi et al, the simulations
are done in 3-D~\cite{Mianroodi2021}.  However, faster diffusivity along 
the dislocation lines is not incorporated in these simulations.
The diffusivity along the core of a dislocation can be 
two to three orders of magnitude higher than the 
bulk diffusivity~\cite{Legros2008}, and hence, 
this can be expected to play a key role in the phase 
transformation mechanism and kinetics.
 
In this paper, we show that faster diffusion 
along the dislocations is indeed essential for 
a system with alloy composition outside spinodal 
limits to show segregation-assisted spinodal 
decomposition along the dislocations. We use a 
phase field dislocation dynamics (PFDD) model. 
There exist several versions of PFDD models in the 
literature: see for example,~\cite{Wang2001,Rodney2003,
Koslowski2002,Shen2004,Lei2011,Hunter2011}. 
Our model is very similar to that of Hunter et al.
~\cite{Hunter2011}. We do not incorporate specific 
crystallographic geometries; however, dislocation-solute
interactions and the segregation effects are 
known to depend primarily on the defect structure and 
their elastic field~\cite{AndersonHirthLothe2017}. In fact, 
we show that our model is able to reproduce the 
experimentally observed: (i) segregation-assisted 
spinodal morphology, and (ii) non-spherical 
precipitate shapes at dislocation intersections, 
in iron based systems, albeit qualitatively.

\section{Formulation and simulation details}
\label{formulation_simulationdetails}

In this paper, we have combined phase field models 
of phase separation~\cite{Hu2001} and dislocation 
dynamics~\cite{Lei2011,Hunter2011}; we incorporate
faster diffusion along the cores of the dislocations 
using the variable mobility methodology proposed by 
Zhu et al.~\cite{Zhu1999}. In this section, since 
all these models are well known, we describe 
the formulation very briefly -- 
for the sake of completeness.

\subsection{Formulation}

We consider a binary system with dislocations; we 
describe the system using two order parameters, 
namely, composition ($c$) and slip ($\eta$). We 
assume the eigenstrain due to the solute 
misfit to be given by: 
\begin{linenomath*}
\begin{equation}
\epsilon^{0c}_{ij} = \epsilon^{*c} \delta_{ij} \beta(c) 
\end{equation}
\end{linenomath*}
where $\epsilon^{*c}$ is the magnitude of the 
eigenstrain,  $\delta_{ij}$ is Kronecker delta 
and $\beta(c)$ is an interpolation function of 
c given by $\beta(c) = c^3(10-15c+6c^2)$. In 
other words, we assume that the solute misfit 
is dilatational and is measured with respect 
to the matrix lattice parameter. The eigenstrain 
due to a dislocation is a pure shear strain~\cite{Mura1987} 
and is obtained as a tensor product of 
the Burgers vector ($\mathbf{b}$) and
the slip plane normal ($\mathbf{n}$) of the 
dislocation; it is given as: 
\begin{linenomath*}
\begin{equation}
\epsilon^{0d}_{ij} = \sum_{\alpha=1}^{N} 
\left(\frac{b_in_j+b_jn_i}{d}\right) \eta_{\alpha}
\end{equation}
\end{linenomath*}
where $d$ is the interplanar spacing, $\alpha$ 
denotes the slip system and $N$ is the total 
number of slip systems.

The stress and strain fields that go into the 
elastic strain energy component are calculated 
by solving the equation of mechanical equilibrium.
We use the Green's function approach \cite{Mura1987} 
to solve the equation of mechanical equilibrium 
-- assuming linear and isotropic elasticity; 
we also assume that the system is elastically
homogeneous.

The total free energy of the system, consists
of, in addition to the elastic component, 
the chemical and the dislocation core energy 
components:
\begin{linenomath*}
\begin{align}
F(c,\eta) = &{N_V}\int \left(Ac^2(1-c)^2 + 
\kappa \left|\nabla c\right|^2 \right)\; 
dV \nonumber \\ &+ \frac{1}{2}\int C_{ijkl}
\epsilon^{el}_{ij}\epsilon^{el}_{kl}\; dV   
+ \sum_{\alpha=1}^N \int B_{\alpha} 
sin^2(\pi\eta_{\alpha})\; dV
\end{align}
\end{linenomath*}
where $N_V$ is the number of atoms per unit 
volume, $V$ is the volume of the system, 
$A$ is a constant that determines the height 
of the bulk free energy density barrier between 
the two phases, $B_{\alpha}$ is the core energy 
coefficient for slip system $\alpha$, $\kappa$ 
is the gradient energy coefficient, $\epsilon^{el}_{ij}$
is the elastic strain (which, in turn, is related 
to the total strain $\epsilon^{t}$ and the 
eigenstrains through the relation 
$\epsilon^{t}_{ij} - \epsilon^{0c}_{ij} 
- \epsilon^{0d}_{ij}$), and $C_{ijkl}$ denotes 
the modulus tensor. Here, we have assumed that 
the interfacial energy between the two phases is 
isotropic.  As can be seen from the total strain 
expression, the $c$ and $\eta$ fields are coupled 
by the elastic strain energy of the system. This 
interaction leads to the segregation of the solute 
atoms to the dislocations. We do not consider 
segregation due to the broken bonds at the core of 
the dislocations in this model. Note that we ignore 
the gradient energy term associated with $\eta$ in 
our formulation. The core width of the dislocations 
in the system is determined by the balance between 
the elastic and the core energies of the dislocation.

Given the free energy functional, the Allen-Cahn 
equation for $\eta$ and the Cahn-Hilliard (CH) 
equation for $c$ can be used to study the 
microstructural evolution in the system. In this 
paper, we are interested in the evolution of the 
composition fields in the presence of the 
dislocations. We assume that the dislocations are 
stationary and hence do not evolve the $\eta$ 
order parameter -- except at the beginning of the 
simulations to equilibrate the dislocation structure. 
Specifically, we use two different dislocation
configurations -- edge dislocation dipoles in 
one and two slip systems. 

The CH equation that described the composition
evolution is given by:
\begin{linenomath*}
\begin{equation}
\frac{\partial c}{\partial t} = \nabla 
\left[ M(\eta) \cdot \nabla \mu^c \right]
\end{equation}
\end{linenomath*}
where $M(\eta)$ is the (scalar) atomic mobility 
which is $\eta$ and hence, position dependent, 
and $\mu^c$ is the variational derivative of 
the free energy per atom with respect to the
composition, and, is given by: 
\begin{linenomath*}
\begin{equation}
\mu^c(c) = 2A(6c^2-6c+1) - 2\kappa \nabla^2 c - 
C_{ijkl}\epsilon^{el}_{kl} \delta_{ij} 
\epsilon^{*c}\beta^\prime(c)
\label{chem-potential}
\end{equation}
\end{linenomath*}
where $\beta^{\prime}(c) = \frac{\partial 
\beta}{\partial c}$. In deriving the above expression, 
we have used the symmetry properties of the elastic 
moduli tensor $C_{ijkl}$.

Let $M_b$ be the bulk mobility of solute 
atoms and let $M_p$ be the enhanced mobility along 
the dislocation core over and above the bulk mobility.
We define the mobility $M$ in the simulation 
domain using the order parameters as follows:
\begin{linenomath*}
\begin{equation}
M(\eta_1,\eta_2) = M_{b} + M_{p}\left\{\mathrm{max}
\left(|\eta_1(1-\eta_1)|, 
|\eta_2(1-\eta_2)|\right)\right\}
\label{mobtwo}
\end{equation}
\end{linenomath*}
where $\eta_1$ and $\eta_2$ are the two order
parameters that describe slip in two different
slip planes, labelled 1 and 2, respectively.
The mobility of atoms given by Eq.~\ref{mobtwo} 
will be highest at points where $\eta_1=0.5$ 
and/or $\eta_2=0.5$. For simulations in which the 
dislocations exist only on a single slip system, 
in this expression, we substitute $\eta_1 = \eta$ 
and $\eta_2 = 0$.

We use a Fourier spectral technique to solve 
the CH equation (which implies periodic 
boundary conditions). The CH equation with the
variable atomic mobility has severe time step 
constraints. Hence, we use the numerical technique 
proposed by Zhu et al.~\cite{Zhu1999}, using which, 
the final evolution equation in the Fourier space 
is given as:
\begin{linenomath*}
\begin{align}
c_{t+\delta t} = c_t + \delta t \left[ \frac{ik 
\lbrace \lbrace M(\eta) \lbrace ik \mu^c(c_n)
\rbrace_F \rbrace_R \rbrace_F }{(1+2\zeta
\delta t\kappa k^4)} \right]
\end{align}
\end{linenomath*}
where $k=\sqrt{k_x^2+k_y^2+k_z^2}$ is the magnitude 
of the reciprocal space vector, the braces denoted 
by $\left\{ \cdot \right\}_R$ and $\left\{ \cdot 
\right\}_F$ indicate the terms in real and Fourier
space respectively, $i=\sqrt{-1}$ and $\delta t$ 
is the time step. As indicated by 
Zhu et al.~\cite{Zhu1999}, the choice of 
$\zeta = \frac{1}{2}\left[\mathrm{max}(M(\eta))
+\mathrm{min}(M(\eta))\right]$ is optimal.

\subsection{Non-dimensionalisation}

All our simulations are carried out using 
non-dimensionalised parameters. Specifically,
we use a characteristic length $L^{\prime}$ of 1\AA, 
a characteristic energy of $E^{\prime} = 0.7\times10^{-21}
\;\mathrm{J}$ and, a characteristic times of $T^{\prime} 
= 200\;\mu\mathrm{s}$ to $T^{\prime} = 200\;\mathrm{s}$.
In dimensional terms, these choices correspond to
a system of size $25.6\;\mathrm{nm}$ for 256 grid 
points, an interfacial energy of  $100\;\mathrm{mJ/m^2}$,
an Young's modulus of $E=130$ GPa and a Poisson's 
ratio of $\nu=0.3$ (isotropic material) and a bulk 
diffusivity of the order of $1\times10^{-16}\;\mathrm{m^2/s}$ 
to $1\times10^{-20}\;\mathrm{m^2/s}$. Finally, the 
used dislocation core energy coefficients imply 
edge dislocations of width $0.7b$ to $5.8b$, where $b$ is the 
magnitude of the Burgers vector. The coherent 
spinodal limits for the parameters chosen by us is 
0.216 and 0.783. Our simulations start with a nominal 
composition well outside the spinodal limit -- 
$c_0=0.11$ to $c_0=0.18$.

The non-dimensionalised parameters and their values 
are listed Table~\ref{tab:simulation_parameters}.
\begin{table}[h]
\centering
\begin{tabular}{c c}
\hline
{\bf Parameter} &  {\bf Value} \\
\hline
Spatial discretisation, $\Delta x=\Delta y=\Delta z$ & 1.0 \\
Time step, $\delta t$ & 0.1 -- 0.5 \\
Elastic moduli: $C_{1111}$ & 175 \\
Elastic moduli: $C_{1122}$ & 75 \\
Elastic moduli: $C_{1212}$ & 50 \\
Bulk free energy coefficient, $A$ & 1 \\
Dislocation core energy coefficient, $B_{\alpha}$ & 0.31 -- 2.52\\
Gradient energy coefficient, $\kappa$ & 1\\
Magnitude of eigenstrain, ${\epsilon^{\star}}^c$ & 0.01 \\
\hline
\end{tabular}
\caption{Non-dimensionalised values of parameters used in
our simulations}
\label{tab:simulation_parameters}
\end{table}
\section{Results and Discussion}
\label{results_discussion}

In this section, we show and discuss segregation 
induced nucleation, spinodal decomposition, 
and concurrent nucleation and spinodal decomposition. 
We rationalise these results in terms of the strength
of the solute-dislocation interaction, the nominal 
alloy composition and the ratio of pipe to bulk mobilities.

\subsection{Segregation induced nucleation}

\begin{figure}[h!]
\begin{center}
\includegraphics[width=11.5 cm]{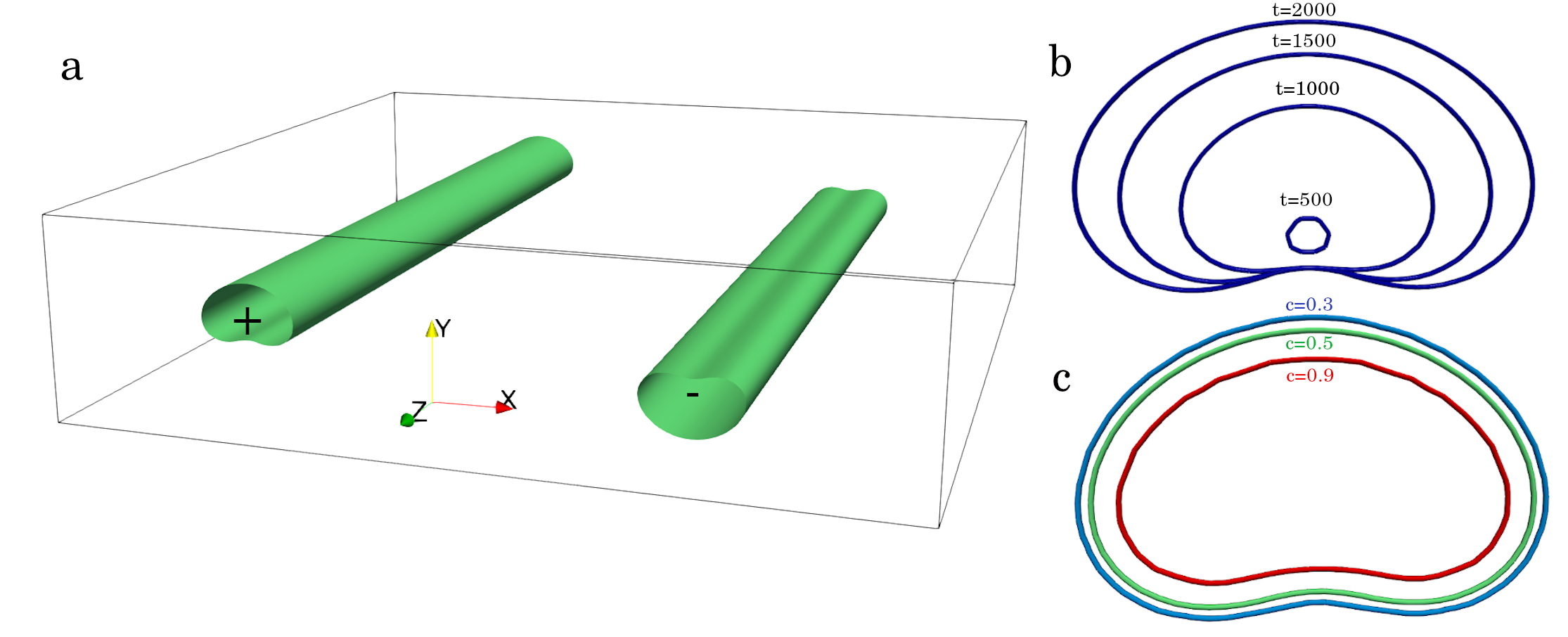}
\caption{\label{Figure1} 
(a) Segregation and nucleation along the dislocations 
in an edge dipole shown using $c=0.5$ isosurfaces aligned 
along the dislocations. (b) shows the $c=0.5$ contours for 
a positive edge dislocation (along the $z$-direction) in 
the $xy$-plane for different time steps, and, (c) shows 
the iso-surfaces for the $c=0.3$, $c=0.5$ and $c=0.9$ 
at t=1500.} 
\end{center}
\end{figure}

In Figure~\ref{Figure1}(a), we show the $c=0.5$ 
isosurface for a system containing an edge 
dislocation dipole. In this and subsequent 
simulations, the edge dislocations are set 128 units 
apart in a system of size $256\times64\times256$;
the nominal alloy composition and the dislocation core
energy coefficient $B_{\alpha}$ are taken as $0.15$
and $0.63$ (unless stated otherwise).

The Burgers vectors of these dislocations are along 
the positive $x$-direction of the simulation cell 
and the slip plane normals for both these dislocations 
are along the positive $y$-direction of the simulation 
cell. These two dislocations are aligned along the $z$-axis
of the simulation cell; the sense vector for the 
dislocation on the left is the positive $z$-direction of 
the simulation cell and the line sense for the dislocation 
on the right is the negative $z$-direction of the simulation 
cell. Note that the strain fields of the positive edge 
dislocation (labelled `+') and the negative edge dislocation 
(labelled `-') are flipped; hence, the accumulation 
of the solute happens above and below the slip plane 
for the positive and negative edge dislocation, 
respectively. Since the results for the two dislocations 
are different only in terms of the sign of the strain fields; 
so, we only consider the positive edge dislocation for the 
rest of this discussion. 
In this simulation, we assume that the bulk and pipe 
mobilities are the same. In Figure \ref{Figure1}(b), 
we show the  $c=0.5$ composition contour lines as a 
function of time in the ($xy$-)plane perpendicular to the 
dislocation line. In Figure \ref{Figure1}(c), we 
show three composition contour lines, namely, $c=0.3$, 
$c=0.5$ and $c=0.9$ in the plane perpendicular to the 
dislocation line.

From these figures, it is clear that there is (a) 
segregation to the dislocation which leads to precipitate 
nucleation and subsequent growth. Further, note 
that the shape of the composition profile is consistent 
with the cardioid profile given 
by~\cite{AndersonHirthLothe2017}:
\begin{linenomath*}
\begin{equation}
c = c_0\;exp\left(\frac{-Ksin\theta}{r}\right) 
\end{equation}
\end{linenomath*}
where $r$ is the radial distance from the dislocation 
line, $c_0$ is the far-field composition, and $K$ is a 
constant. These shapes are also in agreement with 
the shape of the precipitate that nucleates along an 
edge dislocation as shown in~\cite{WolferAshkin1976,Hu2001}. 
Thus, when we do not assume faster pipe diffusion, we 
recover the results in the existing literature, which, 
in turn, serves as a benchmark for our model and implementation.

\subsection{Spinodal decomposition induced by enhanced pipe diffusion}

\begin{figure}[h!]
\begin{center}
\includegraphics[width=0.85\textwidth]{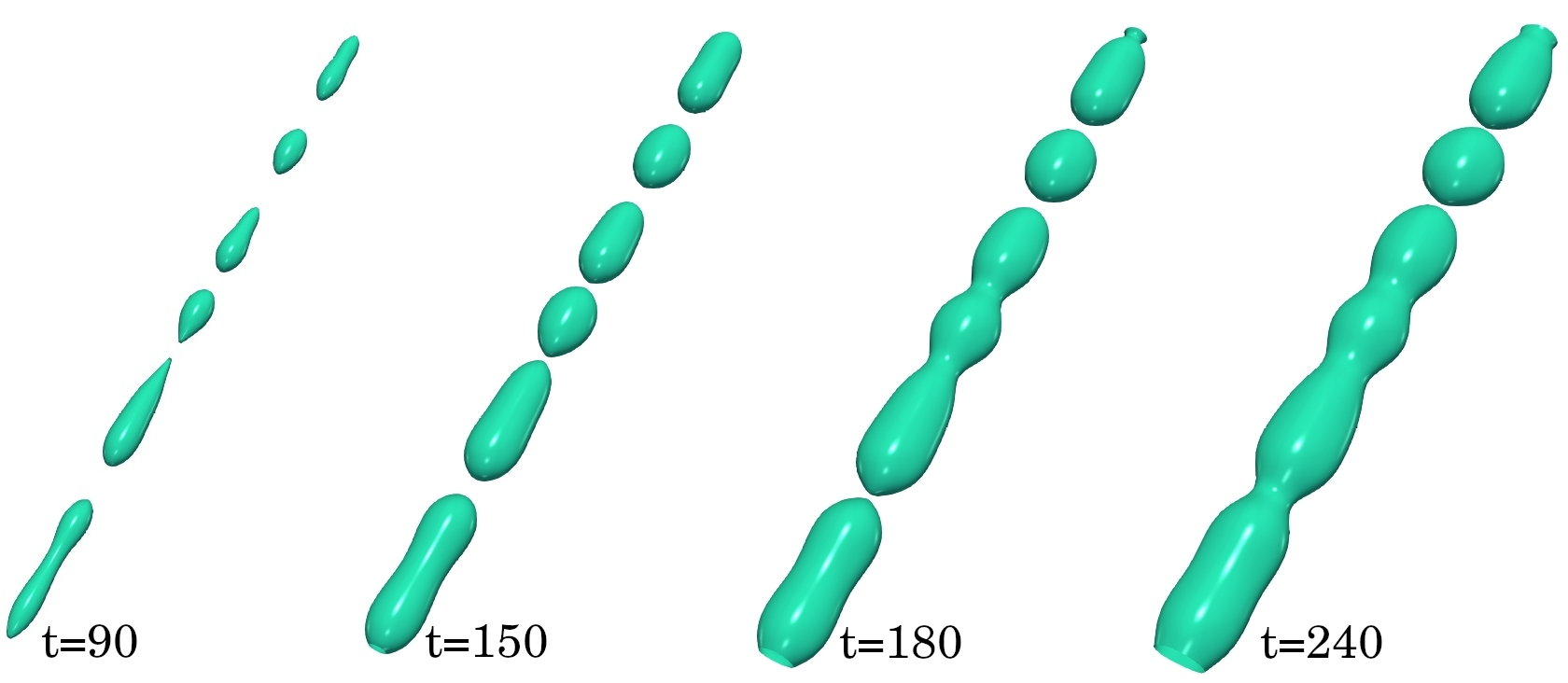}
\caption{\label{Figure2} The $c=0.5$ contours along the 
dislocation line for a system in which dislocation 
core mobility is one hundred times 
that of the bulk. In this figure, we show the 
evolution of composition only along the positive
edge dislocation. The spinodal fluctuations 
along the dislocation line grow due to the higher 
diffusivity along the dislocation core. The phase 
separated regions in the form of blobs coalesce as 
shown in (c) and (d) creating a linearly elongated
morphology along the dislocation line.} 
\end{center}
\end{figure}

\begin{figure}
\centering
\subfigure[]{
\centering
\includegraphics[width=0.4\textwidth,trim=0.2cm 0.2cm 0.2cm 0.2cm,clip=true]{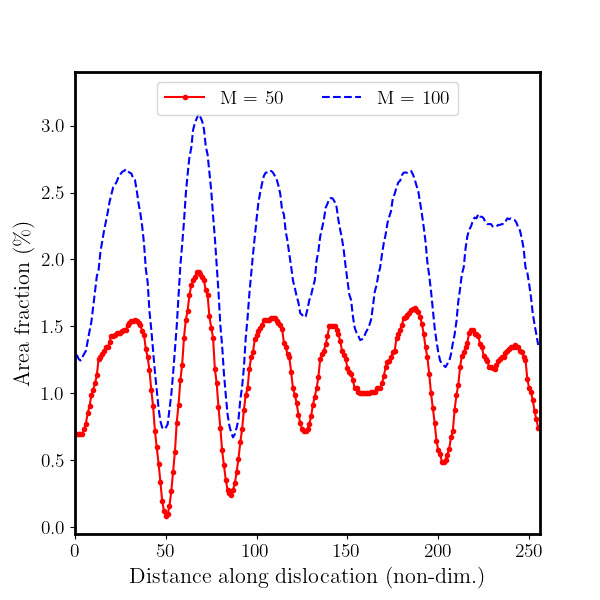}}
\subfigure[]{
\centering
\includegraphics[width=0.4\textwidth,trim=0.2cm 0.2cm 0.2cm 0.2cm,clip=true]{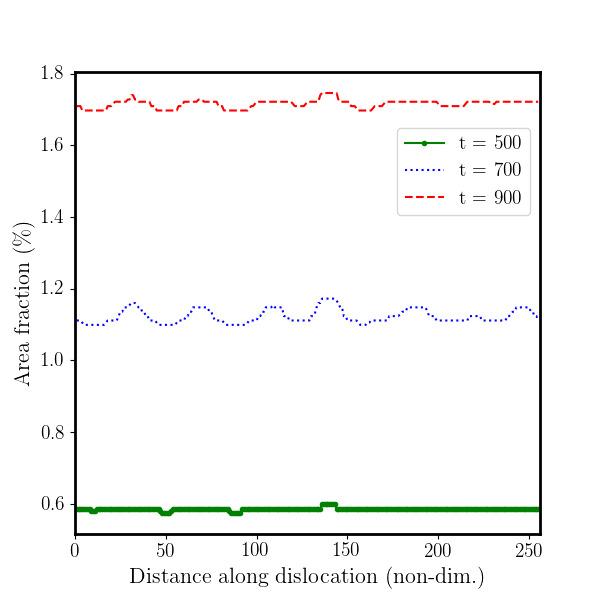}} 
\caption{\label{Figure3} (a) Area fractions of the 
solute-rich phase formed near the dislocation line 
at $t=385$ for $\frac{M_p}{M_b}=50$ and 
$\frac{M_p}{M_b}=100$, (b) Area fraction 
at three different time steps 
for $\frac{M_p}{M_b}=10$}
\end{figure}

Let us consider the same initial profile as in the 
previous section; however, let us increase the pipe 
mobility to be hundred times that of the bulk mobility. 
In Figure \ref{Figure2}, we show the $c=0.5$ composition
contour lines for this system. It is clear that the 
system undergoes phase separation via the spinodal mechanism 
along the dislocation line. As time progresses, there is 
also coarsening and coalescence. Thus, in contrast to 
Figure~\ref{Figure1} where we saw cardioid shape, here, 
we see a morphologically discontinuous phase separated 
microstructure along the dislocation line at earlier times
which becomes a nearly continuous morphology albeit
with a wavy interface.
In order to see the effect of the ratio of the pipe mobility 
to bulk mobility, we also carried out simulations where the 
atomic mobility was kept as ten and fifty times that of 
bulk mobility. These results are summarised in 
Figure~\ref{Figure3} by plotting the area fraction in the 
left half of the simulation cell which contains the positive 
edge dislocation. We calculate the area fraction (in 
percentage -- in the $xy$-plane) of the solute-rich phase 
along the dislocation line by counting the number of grid 
points with $c\geq 0.5$ as a fraction of the total number 
of grid points in the left half of the $xy$-plane. 
In Figure \ref{Figure3}(a), we show the area fraction 
(which is a proxy for the morphology of the solute rich phase) 
along the dislocation line for the cases where the atomic 
mobility along the dislocation core is 100 and 50 times 
that of the bulk -- for two different times, 
namely $t=200$ and $t=385$. Both $M=50$ and $M=100$ cases 
show wavy profiles, indicating the growth of the composition 
fluctuation by spinodal mechanism. However, the amplitudes 
are different; smaller the pipe mobility, smaller the amplitude 
of the profile; in fact, the profiles, but for the amplitude 
are almost the same. Since the bulk mobility is 
the same in both cases, the amount of solute reaching the dislocation
at any given time is the same. Thus, the enhanced amplitude of 
fluctuation is solely due to the higher pipe mobility. 

From the classical theory of spinodal decomposition, 
it is known that the coherent spinodal limit can be 
obtained by solving the following equation \cite{Cahn1962}:
\begin{linenomath*}
\begin{equation}
2A(6c^2-6c+1) + \frac{2(\epsilon^{*c})^2E}{1-\nu} = 0
\label{critlambda}
\end{equation}
\end{linenomath*}
For our system, the coherent spinodal limit according
to Eq. (\ref{critlambda}) is calculated as $c_{crit} = 
0.216$. Even for our initial composition of 
$c=0.15\pm0.005$, which is well outside this spinodal 
limit, we observe that the solute-dislocation interaction 
is strong enough to segregate enough material along the 
dislocation and increase the local composition field above
the coherent spinodal limit. As the composition along the 
dislocation enters the spinodal regime, there are also 
the so-called ``up-hill'' diffusion fluxes set-up along 
the dislocation line.

From the classical theory of spinodal decomposition,
it is also known that, at the early stages of the 
decomposition, at least, the rate of growth of the 
amplitude of a fluctuation of wavelength $\lambda$ 
is given by$A(\lambda,t) = A(\lambda,0)
\exp[R(\lambda)t]$~\cite{Cahn1962} where $R(\lambda)$ 
is the amplification factor. This amplification 
factor is a function of mobility $R(\lambda) = 
-M\lambda^2\left[\frac{\partial^2 g(c)}
{\partial c^2} + 2 (\epsilon^{*c})^2 E + 
2\kappa \lambda^2 \right]$, where M is the mobility, 
and $g(c)$ is the bulk free energy density given 
by the double well potential. This expression 
assumes that the system under consideration is 
elastically isotropic. The composition fluctuation 
decays when $R(\lambda)$ is negative and
grows when it is positive. Since the amplification 
factor is a function of the wavelength of the fluctuation, 
there is a critical wavelength $\lambda_c$ at which 
$R(\lambda)$ will be equal to zero. The critical 
wavelength $\lambda_{c} = 
\left\{-\frac{8\pi^2\kappa}{\left[ \frac{\partial^2 g(c)}
{\partial c^2} + 2 (\epsilon^{*c})^2 E \right]} 
\right\}^{\frac{1}{2}}$ and the maximally growing 
wavelength $\lambda_{max} = \frac{\lambda_c}
{\sqrt{2}}$. That is, the energetic 
parameters of the system such as the (incipient) 
interfacial energy, the eigenstrains, and the elastic 
moduli determine the wavelengths of compositional 
fluctuations; on the other hand, the growth rate 
depends on the mobility. Thus, in 
Figure~\ref{Figure3}(a), the mobilities do not 
change the profile and only change their amplitude is 
consistent with the classical theory 
of spinodal decomposition.

In Figure \ref{Figure3}(b), we show the profile of 
the area fraction plot at different times for the case 
where the atomic mobility along the dislocation line is 
10 times that of bulk mobility. Here, we see that with 
time, the morphology becomes more uniform and the 
waveiness is smoothened out. This can be understood 
in terms of the stability of the compositional cylinder 
(of cardioid cross-section) along the dislocation line. 
As shown by Nichols and Mullins~\cite{Nichols1965}, the 
morphological perturbations along the length of the 
cylinder (albeit, of circular cross-section) decay to 
material flow from the bulk. Thus, the spinodal fluctuations 
along the dislocation line, which tend to perturb the 
cylinder, will die, unless, the spinodal fluctuations grow 
faster than the rate at which bulk diffusion is trying to 
smoothen these morphologies. This is the reason why we 
see spinodal morphology only with higher mobilities.   

Thus, we see that in the case of dislocation assisted
phase separation, the mechanism of phase separation
can change from nucleation and growth to spinodal 
decomposition depending on the enhanced pipe mobility. 
Given that the pipe mobilities are typically two to three
orders of magnitude higher, in systems that undergo
spinodal decomposition, the dislocations can play a crucial
role in the phase separation mechanism; more importantly,
dislocations can promote phase separation even when the
nominal alloy composition is outside the spinodal region.
Thus, the spinodal phase separation in these systems 
is induced by segregation and is assisted by pipe diffusion.

\subsection{Effect of dislocation network: concurrent
nucleation and spinodal}

\begin{figure}[h!]
\begin{center}
\includegraphics[width=0.85\textwidth]{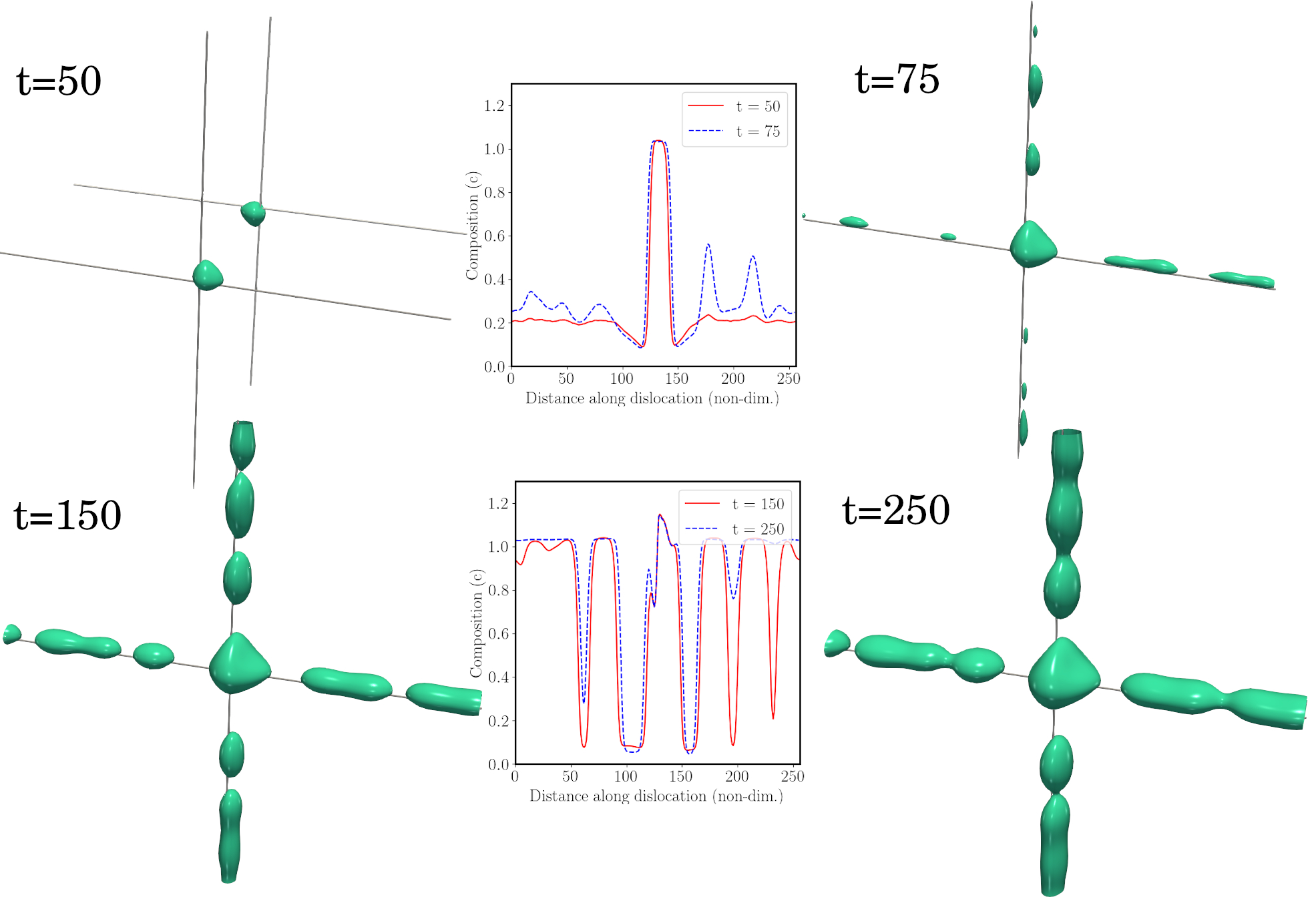}
\caption{\label{Figure4} Growth of spinodal fluctuations 
along two intersecting dislocations. The horizontal and 
vertical lines are drawn through the two dislocation 
cores, as a guide to the eye. The vertical and horizontal 
lines are parallel to the y and z axes respectively.  
At t=50, both pairs of dislocation intersections are shown 
for better understanding. The composition profile 
along the z direction through the dislocation core is 
shown in the inset.} 
\end{center}
\end{figure}

Our aim, in this subsection, is to  understand the 
effect of dislocation networks. So, let us consider 
a system with two intersecting slip systems; in addition to the 
dipoles in the (010) plane (as in the previous cases), 
we introduce another dipole in the (001) plane 
(that is, with the slip plane normals in the positive 
$z$-direction) with the sense vectors of the dislocations 
along the positive and negative $y$-direction (for the 
dislocations at the left and right, respectively). 
In this case, we use $256\times256\times256$ grid points
and the Burgers vectors are aligned along the 
$x$-direction. This configuration is shown in 
Figure~\ref{Figure4}(a). In Figure~\ref{Figure4}(a)-(d) 
we show the $c=0.5$ iso-surfaces at different times; 
in the figures (b)-(d), we show only one intersection 
for clarity. In this case also, the pipe mobility is 
100 times the bulk mobility.
The initial overall composition in the simulation cell 
is $c=0.15\pm0.005$. At $t=50$, there solute-rich phase 
nucleates at the intersection of the two dislocations.
As time progresses, the composition along the other parts
of the two intersecting dislocations cross the spinodal
limit of $c^{crit} = 0.216$. At $t=75$, there is
phase separation along the dislocation lines via 
spinodal mechanism, and subsequently, there
is growth and coalescence. 

In order to clearly show the concurrent
nucleation and growth at the junction, and spinodal
along the dislocations, we also show the composition 
profile along one of the dislocations in the inset. 
The composition profile in the initial stages clearly 
shows the nucleation of the second phase at 
the intersection. The composition along the rest of 
the dislocation reaches the coherent spinodal limit as time 
progresses, and, at $t=75$, the composition along 
the dislocation line is above the spinodal limit; 
hence, we observe growing composition fluctuations 
on both sides of the central precipitate. These 
fluctuations along the dislocation lead to solute 
rich second phase. When phase separation occurs, 
material is also added to the central precipitate 
by virtue of the faster atomic mobility along the 
dislocation core. In the composition profiles in 
Figure~\ref{Figure4}, the values above unity
are due to Gibbs-Thomson effect. 

\subsection{Map of phase separation mechanisms} 
\label{parameters_discussion}
\begin{figure}[h!]
\centering
\subfigure[]{
\centering
\includegraphics[width=0.4\textwidth]{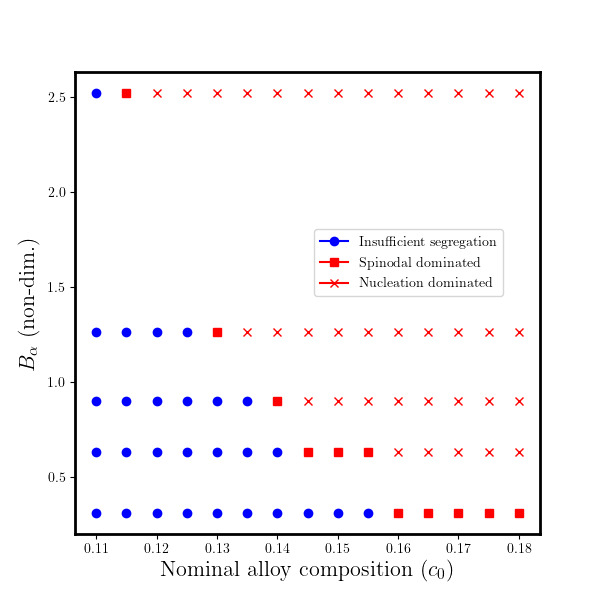}}
\subfigure[]{
\centering
\includegraphics[width=0.4\textwidth]{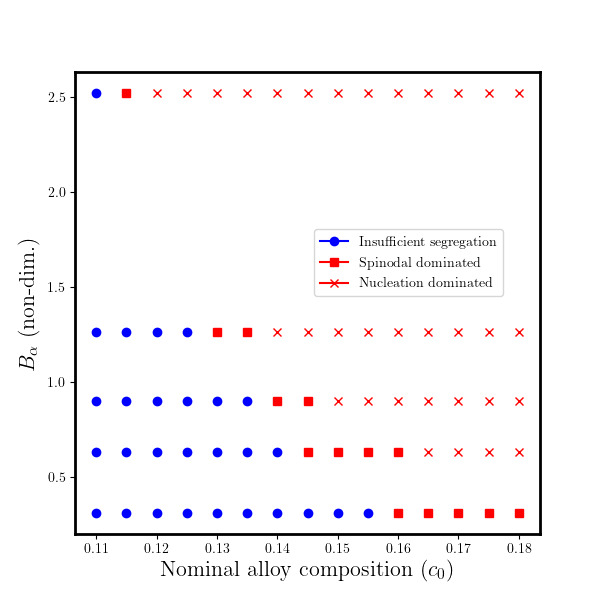}}	
\caption{ \label{Figure5} The occurrence of spinodal 
dominated and segregation dominated morphologies 
are shown for different far field compositions 
($c_0$) and core energy coefficients ($B_{\alpha}$) 
at a ratio of mobilities  of (a) $\frac{M_p}{M_b}
=100$ and (b) $\frac{M_p}{M_b}=300$.}
\end{figure} 
From the results and discussion so far, 
it is clear that the segregation (which is 
controlled, in our model, by the core energy 
coefficient), the available solute (which is 
determined by the overall alloy composition), 
and the pipe mobility. Figure \ref{Figure5} shows 
a map of phase separation mechanisms as a function 
of the core energy coefficient $B_{\alpha}$, overall 
alloy composition $c_0$ and mobility ratio 
$\frac{M_p}{M_b}$. We considered two different pipe 
mobilities of $\frac{M_p}{M_b}=100$ (Figure \ref{Figure5}(a)), 
and, $\frac{M_p}{M_b}=300$ (Figure \ref{Figure5}(b)). 
For increasing $B_{\alpha}$, the core width decreases, 
causing higher segregation of solute to dislocation by 
elastic interaction. Due to this, segregation 
assisted phase transformation occurs at smaller 
$c_0$ for higher $B_{\alpha}$. Phase separation occurs 
only in those cases where the perturbations 
caused by the pipe mobility grows at a faster 
rate as compared to material segregating from 
the bulk~\cite{Nichols1965}. This is clear from 
Figure \ref{Figure5}(b), in which there is a wider range 
of $c_0$ across which spinodal morphology dominates. 

\section{Comparison with experimental results}
\label{compare_experiments}

Even though a direct comparison of our simulation
results with experiments is not possible, a qualitative
comparison is possible. In this section, we make two
such comparisons. 

\subsection{Dislocation induced spinodal}
\begin{figure}[h!]
\centering
\subfigure[]{
\centering
\includegraphics[width=0.23\textwidth]{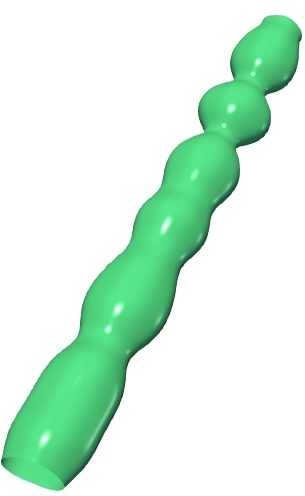}}
\subfigure[]{
\centering
\includegraphics[width=0.24\textwidth]{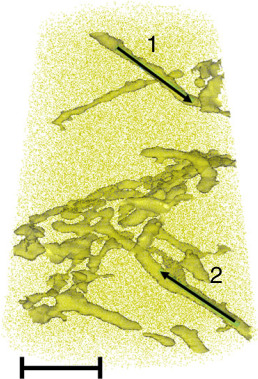}}
\subfigure[]{
\centering
\includegraphics[width=0.24\textwidth,trim=0.0cm 0.0cm 0.0cm 0.6cm,clip=true]{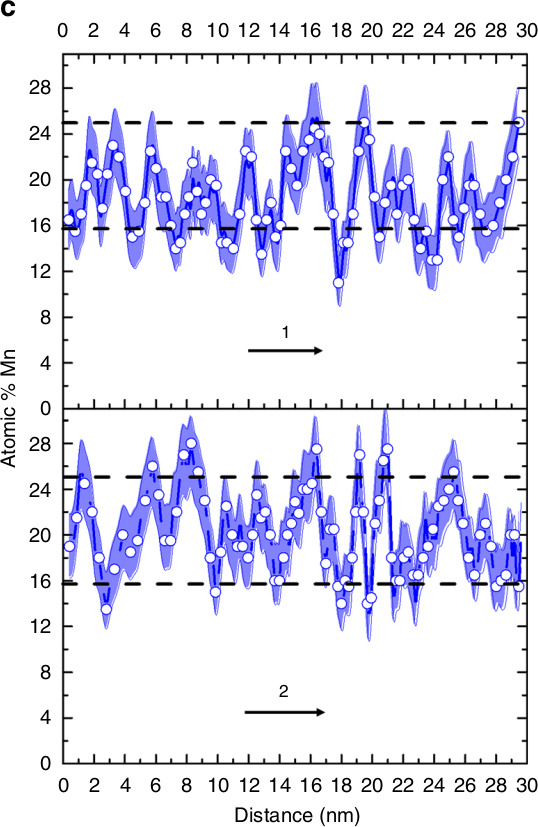}}	
\caption{\label{Figure6} Comparison of (a) the morphology formed 
by localized spinodal decomposition in our simulation 
(with pipe-diffusion) ($t=385$) with (b) the atom probe 
results of~\cite{DaSilva2018} (scale bar represents a 
length of 40 nm). The morphology along the arrows 
indicated in (b) match well with the output from our 
simulations. (c) shows the atomic fraction of Mn plotted 
along the dislocation line. The dashed lines show the 
typical fluctuation values. The sub-figures (b) and (c) 
are cropped from Figure 1 of Da Silva et al~\cite{DaSilva2018} with the
labels removed from the cropped area, and is 
shared under Creative Commons Attribution 4.0 International License:
{\tt http://creativecommons.org/licenses/by/4.0/}.}
\end{figure} 
In Figure~\ref{Figure6}, we show (a) the
spinodally decomposed morphology along the 
dislocation line from our simulations; in (b) and (c), 
the atom probe results of localised spinodal flcutuations 
in the Fe-Mn system in which the overall alloy 
composition (9 at.\% Mn) was outside the spinodal 
limits (15.8--25 at.\% Mn) at 450 \textdegree C~\cite{DaSilva2018}. 
As can be seen, there are composition fluctuations measured 
along the dislocations in these experiments.
The morphology of phase separated regions seen 
in our simulations is very similar to that observed 
in these atom probe results.

Further, our results clearly show the role of faster 
diffusion along the dislocation cores in producing 
such morphologies; as shown in the simulations and 
explained using the linear stability analysis of 
Nichols and Mullins~\cite{Nichols1965}, if the bulk 
and dislocation pipe mobilities are assumed to be the same, 
this phenomenon will not be observed; this, we believe 
could indeed be the case with some of the previous simulation 
studies~\cite{Mianroodi2021}.

\subsection{Effect of dislocation network on precipitate
morphology}

A precipitate that nucleates homogeneously in an 
isotropic system will grow with a spherical shape. 
However, the presence of elastic fields of defects 
in the system can alter this behaviour and can 
affect the shape of the precipitate. 

A recent APT study in maraging steel~\cite{Kevin2021} 
has shown that precipitates formed in high dislocation 
density regions have a markedly different morphology as
compared to the defect-free region. The Fe-Mo preipitates 
assume a near spherical shape in defect free regions and 
deviate from this shape as the precipitation happens 
in regions with dislocation networks.  
\begin{figure}
\centering
\subfigure[]{
\centering
\includegraphics[width=0.15\textwidth]{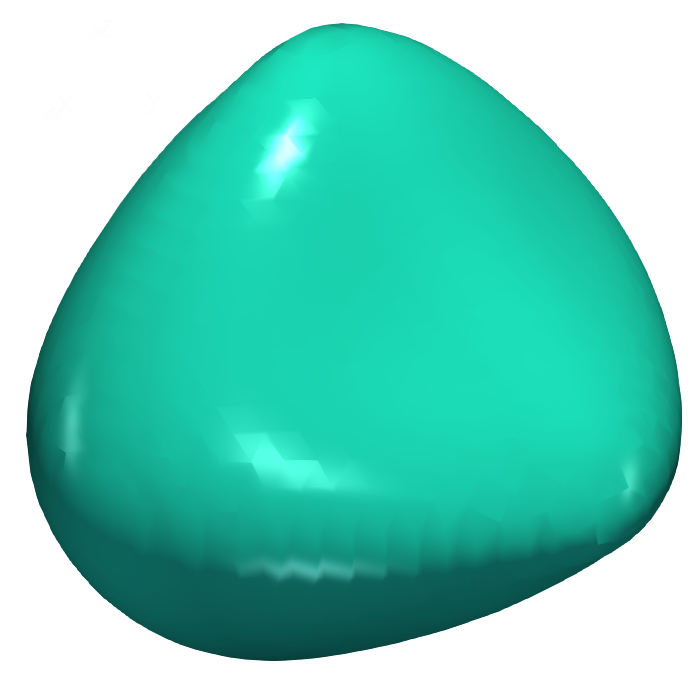}}
\subfigure[]{
\centering
\includegraphics[width=0.15\textwidth]{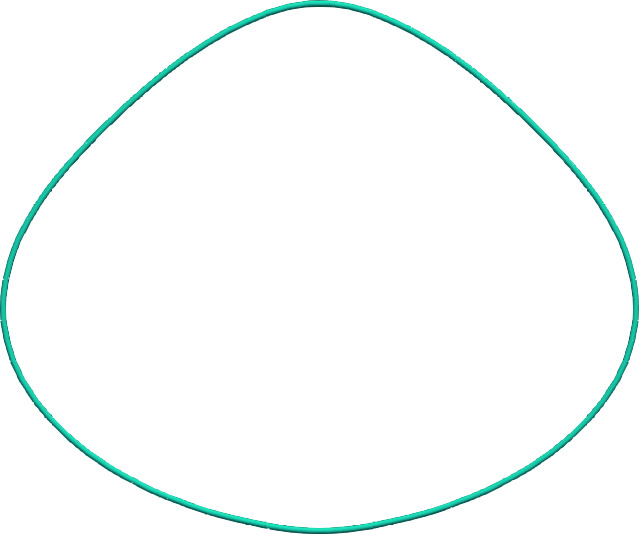}} 
\subfigure[]{
\centering
\includegraphics[width=0.15\textwidth]{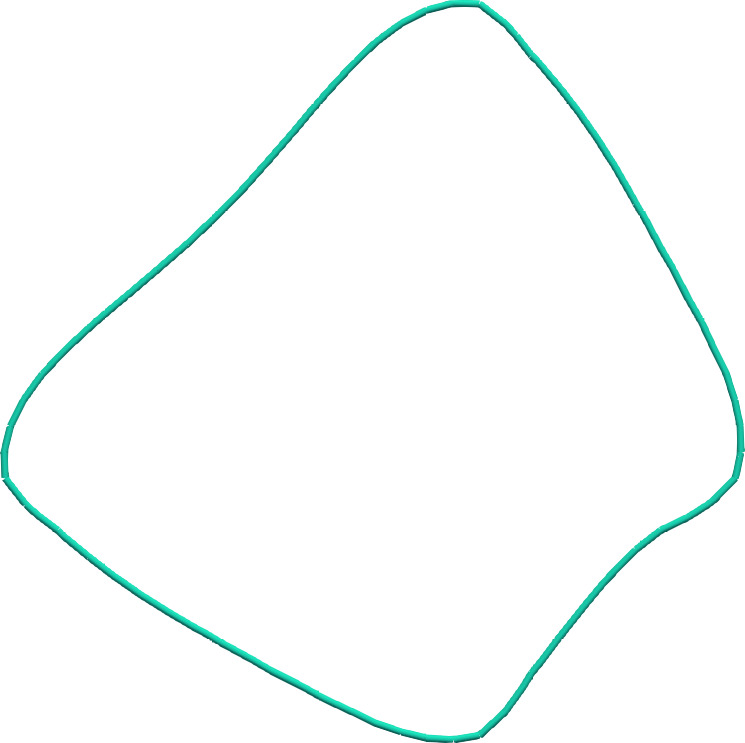}}
\subfigure[]{
\centering
\includegraphics[width=0.15\textwidth]{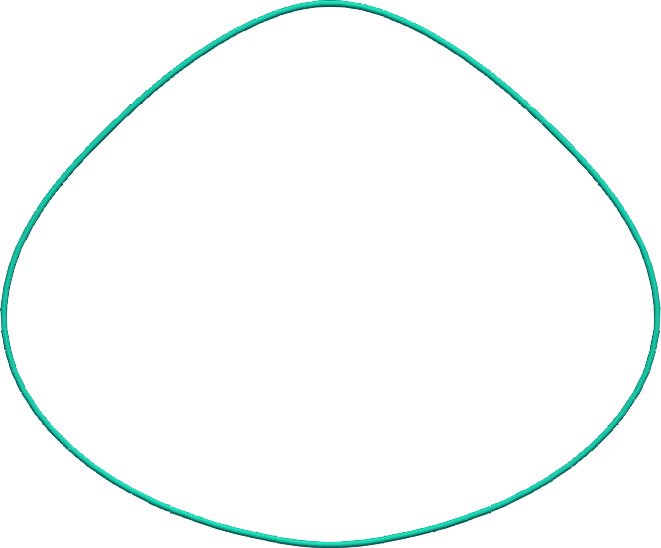}}
\subfigure[]{
\centering
\includegraphics[width=0.3\textwidth]{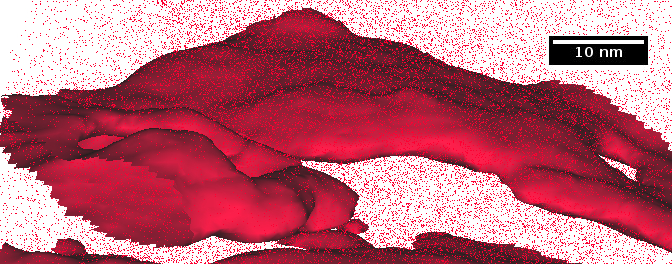}}
\caption{(a) shows the morphology of the precipitate 
formed at the intersection of two dislocations in our 
simulation. (b), (c) and (d) show the section contours of 
the precipitate along the $xy$, $yz$ and $zx$ planes, 
respectively. (e) shows the Mo iso-surfaces (11 at. \%) 
depicting a precipitate in a Maraging steel, that nucleates 
in a region with high dislocation density 
(see \cite{Kevin2021})}
		\label{Figure7}
\end{figure} 
In Figure~\ref{Figure7}(a), we show the morphology 
of the precipitate that nucleates at the dislocation 
intersection; as can be seen in Figure~\ref{Figure7}(b)-(d), 
the cross-sections along different planes are very different 
and the precipitate is highly non-spherical. In fact, we 
have verified that depending on the nature of the 
dislocations that intersect (positive-positive, 
negative-positive, positive-negative, and negative-negative), 
the cross-sections differ (see supplementary information). 
Thus, the protuberances in the precipitate shape are a result 
of the interaction of solutes with dislocation strain fields. 
Very similar morphologies are seen by 
Jacob et al.~\cite{Kevin2021} in their APT studies  of 
Fe-Mo precipitates: see Figure~\ref{Figure7}(e). While the 
protuberances in the experimental morphologies are very 
similar, we believe that the plate-like structure of 
the precipitate is a result of the intersection of a 
large number of dislocations.

\section{Conclusions}
\label{conclusions}

The following are the salient conclusions from our 
phase field simulations of segregation assisted and 
pipe-diffusion driven spinodal decomposition along 
dislocations.  
\begin{itemize}
\item For a system with alloy composition well 
outside the spinodal limit, when there is faster 
mobility at the dislocation core, phase transformation 
occurs by spinodal decomposition along the dislocation 
line. The rate of growth of composition fluctuations 
increases with increasing pipe mobility. There is 
qualitative agreement between the spinodal morphology 
along the dislocation line obtained in our simulations 
with that of APT experiments for Fe-Mn 
systems~\cite{DaSilva2018}.
\item We have identified the range of parameters, namely, 
dislocation-solute interaction strength 
(determined by $B_{\alpha}$), nominal alloy 
composition and the enhanced pipe mobility, a combination of which 
causes phase separation by spinodal decomposition.
\item Concurrent nucleation and spinodal decomposition 
is possible in the case of systems with dislocation 
networks, where the intersections of dislocations act 
as energetically favourable sites for nucleation. The 
morphology of the precipitate at the dislocation 
intersection agrees well with the Mo-rich precipitates 
in maraging steels~\cite{Kevin2021}.   
\end{itemize}

\section*{Acknolwedgements}

We thank (i) Mr Kevin Jacob, and, Professor Nagamani 
Jaya Balila for sharing their experimental results 
(Figure~\ref{Figure7}(e)) and for useful discussions; 
(ii) Professor Ferdinand Haider, Professor T A Abinandanan, 
and Mr. Abhinav Roy  for discussions on various aspects 
of the formulation and numerical implementation; and, 
(iii)  (a) Dendrite and Space-Time, IIT Bombay, (b) Spinode 
– the DST-FIST HPC facility, Dept. of ME \& MS, IIT Bombay, 
and (c) C-DAC, Pune for high performance computing 
facilities. This project is funded by Science and Engineering 
Research Board, Dept. of Science and Technology, 
Govt. of India though the research grant CRG/2019/005060.

\bibliography{mybibfile}

\pagebreak

\section*{Supplementary information}
\begin{figure}[h!]
\begin{center}
\includegraphics[width=0.8\textwidth]{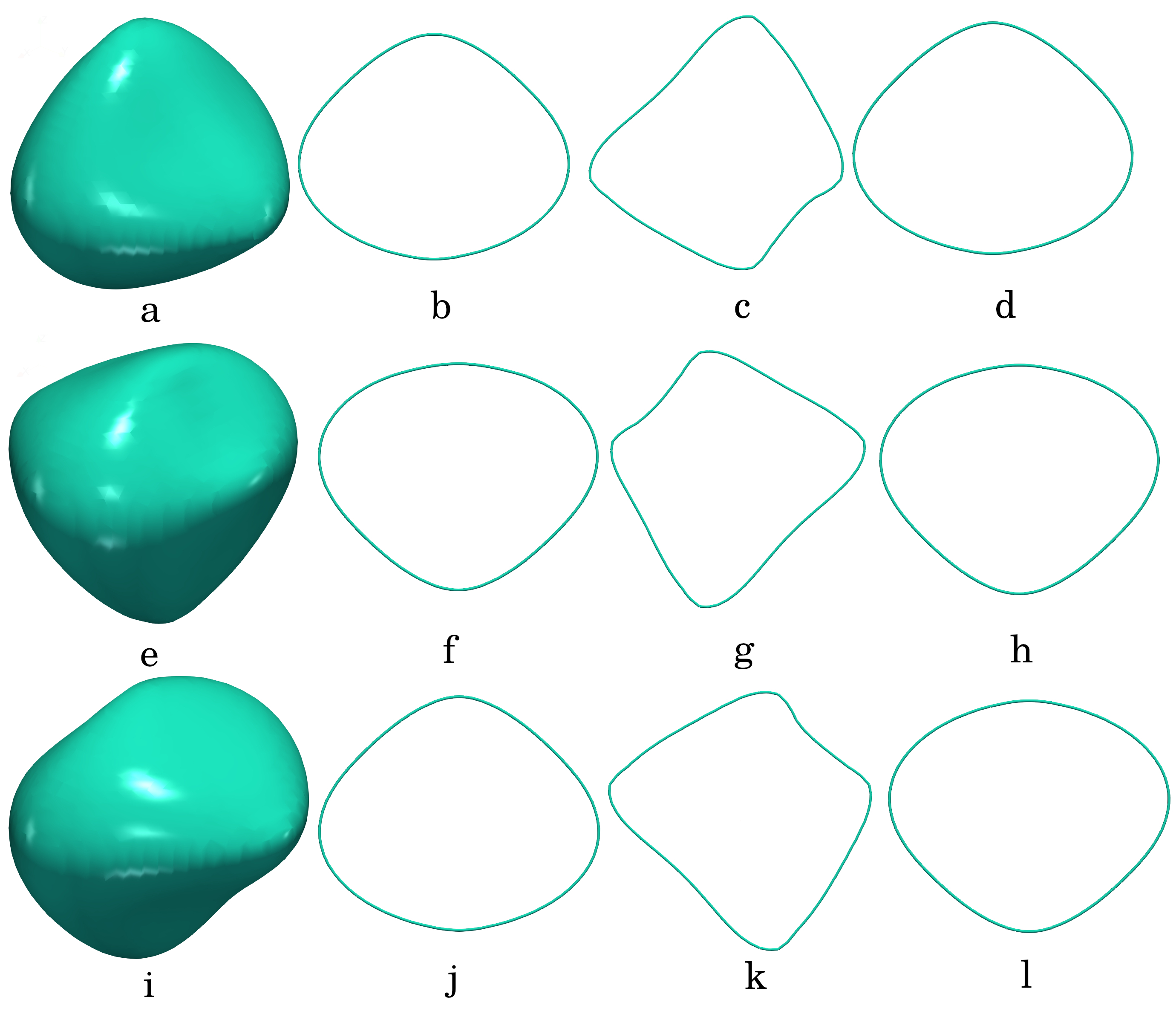}
\caption{\label{FigureS1} The morphology of the precipitate 
formed at the intersection of dislocations for 
(i) positive-positive (a-d), (ii) negative-negative (e-h) 
and (iii) positive-negative (i-l) dislocation combinations. 
The figures (b), (f) and (j) show the $xy$ section; (c), 
(g) and (k) show the $yz$ section; and (d), (h) and (l) show 
the $yz$ section of the precipitate. } 
\end{center}
\end{figure}
\end{document}